\documentclass[article,preprint, onecolumn,nofootinbib,prd,superscriptaddress]{revtex4}
\usepackage[letterpaper,left=.75in,right=.75in,top=.75in,bottom=1.in]{geometry}
\usepackage{graphicx,amsfonts,color,comment,amsmath,hyperref,float}
\usepackage{amssymb}	
\usepackage{scrextend}

\usepackage{mathrsfs,amssymb}  
\usepackage{cancel}
\usepackage[normalem]{ulem}

\newcommand{\be}{\begin{equation}}
\newcommand{\ee}{\end{equation}}
\newcommand{\bea}{\begin{eqnarray}}
\newcommand{\eea}{\end{eqnarray}}
\def\tb{t_{\beta}}

\def\sb  {s_{\beta}}
\def\cb  {c_{\beta}}
\def\stwob  {s_{2\beta}}
\def\ctwob  {c_{2\beta}}
\def\sa  {s_{\alpha}}
\def\ca  {c_{\alpha}}

\def\sba  {s_{\beta-\alpha}}
\def\cba  {c_{\beta-\alpha}}
\def\tanb{\tan\beta}
\def\cotb{\cot\beta}

\def\lam{\lambda}

\def\lambar{\lam}

\def\cbma{c_{\beta-\alpha}}

\def\ctwob{c_{2\beta}}
\def\stwob{s_{2\beta}}
\def\sthreeb{s_{3\beta}}

\def\cthreeb{c_{3\beta}}

\def\lamhat{\widehat\lam}

\def\ha{A}
\def\mha{m_{\ha}}
\def\nicefrac#1#2{\hbox{${#1\over #2}$}}

\begin{document}

\title{Asymmetric Dark Matter Models and the LHC Diphoton Excess}

\author{Mads T. Frandsen}
\affiliation{CP$^{3}$-Origins \& Danish Institute for Advanced Study {DIAS}, University of Southern Denmark, Campusvej 55, DK-5230 Odense M, Denmark}
\author{Ian M. Shoemaker}
\affiliation{Department of Physics; Department of Astronomy \& Astrophysics; Center for Particle and Gravitational Astrophysics, The Pennsylvania State University, University Park, PA 16802, USA}

\preprint{CP3-Origins-2016-14 DNRF90 and DIAS-2016-14}
\date{\today}
\begin{abstract}
{The existence of dark matter (DM) and the origin of the baryon asymmetry are persistent indications that the SM is incomplete. More recently, the ATLAS and CMS experiments have observed an excess of diphoton events with invariant mass of about 750 GeV. One interpretation of this excess is decays of a new spin-0 particle with a sizable diphoton partial width, e.g. induced by new heavy weakly charged particles. These are also key ingredients in models cogenerating asymmetric DM and baryons via sphaleron interactions and an initial particle asymmetry.  We explore what consequences the new scalar may have for models of asymmetric DM that attempt to account for the similarity of the dark and visible matter abundances. }
\end{abstract}


\maketitle

\section{Introduction}

In  \cite{ATLAS,CMS:2015dxe} the ATLAS  and CMS experiments both report excesses in diphoton final states with invariant masses around 750 GeV. If these excesses are interpreted as a narrow resonance the local significances correspond to 3.6 and 2.6 $\sigma$ respectively leading to modest global significances of 2$\sigma$ in ATLAS and 1.2$\sigma$ in CMS --- The significance is slightly higher in the ATLAS data if interpreted as a resonance with a width of about 50 GeV while the CMS significance is decreased slightly. Preliminary updates presented at Moriond 2016 show a slight increase in the significance of the CMS results with more data, while the ATLAS  reanalysis of the 8 TeV data, results in a 2 sigma excess in the same mass region and thus in less tension between the 13 and 8 TeV data. 

Although the significances of the results are modest it is tantalizing to interpret this as the first glimpse of new physics beyond the SM. It is then particularly motivated to investigate how this putative resonance may fit into a bigger picture of the origin of mass, i.e. electroweak symmetry breaking or the origin of dark matter (DM). Here we focus on a possible connection with the origin of DM, motivated by the observation that the same basic ingredients, a new scalar resonance and new weak charged states, can account for the diphoton excess and are required in models of asymmetric DM.

\bigskip
One interpretation of the diphoton excess is the $s$-channel production, via gluon (and photon) fusion, of a new (pseudo) scalar resonance $\phi$ with a mass $m_\phi \sim 750$ GeV 
\begin{align}
gg\,  (\gamma\gamma) \to \phi \to 2\gamma Ê, 
\end{align}
 and a cross-section into diphotons of about  $\sigma_{ \gamma\gamma} \sim 5-10~{\rm fb}$. 
The required cross-section, and the absence of signals in other decay channels, indicates the presence of new (weak) charged states to enhance the diphoton decay width $\Gamma_{\phi\to \gamma\gamma}$. Alternatively new colored states could enhance both the production cross-section and the partial width into diphotons, but we do not consider new colored states in this study. Importantly, gluon induced production grows by a factor of $\sim 5$ between 8 TeV and 13 TeV collisions, improving agreement between the putative 13 TeV signal and the 8 TeV data over e.g. $q\bar{q}$ induced production as has been discussed in several studies, e.g.~\cite{Franceschini:2015kwy}. The diphoton induced production and its scaling between 8 TeV and 13 TeV is more uncertain but has been argued to grow significantly from 8 TeV to 13 TeV, see e.g.
~\cite{Fichet:2015vvy,Csaki:2015vek,Sahin:2016lda,Fichet:2016pvq,Harland-Lang:2016apc,Molinaro:2015cwg}.

 \bigskip
 Meanwhile the cosmological dark matter abundance is curiously close to the abundance of baryons, suggesting a common origin. Since the baryonic density is known to originate from a primordial particle-antiparticle asymmetry, it is natural to consider the possibility that DM emerged from a related mechanism. Asymmetric DM provides one compelling framework of this. 
Models for the cogeneration of a DM and baryon asymmetry, able to explain the observed relic density of DM today, require the same new states as may explain the diphoton excess --- i.e. new electroweak charged states, coupled to the EW sphalerons and new scalar states mediating the decay of these EW charged states into neutral DM particles, as in e.g  \cite{Barr:1991qn,Frandsen:2011kt,Perez:2013tea}. {Already there has been considerable work on the possible connections of the diphoton resonance to DM}~\cite{Franceschini:2015kwy,Knapen:2015dap,Dev:2015isx,Martinez:2015kmn,Bauer:2015boy,Mambrini:2015wyu,Han:2015cty,Han:2015yjk,Ghorbani:2016jdq,Berlin:2016hqw,Okada:2016rav,Cao:2016cok,Ge:2016xcq,Tsai:2016lfg,D'Eramo:2016mgv,Chen:2016sck,Belanger:2016ywb,DeRomeri:2016xpb,Arcadi:2016dbl,Morgante:2016cfv}.

 \begin{figure}[t]
\begin{center}
\includegraphics[width=.65\textwidth]{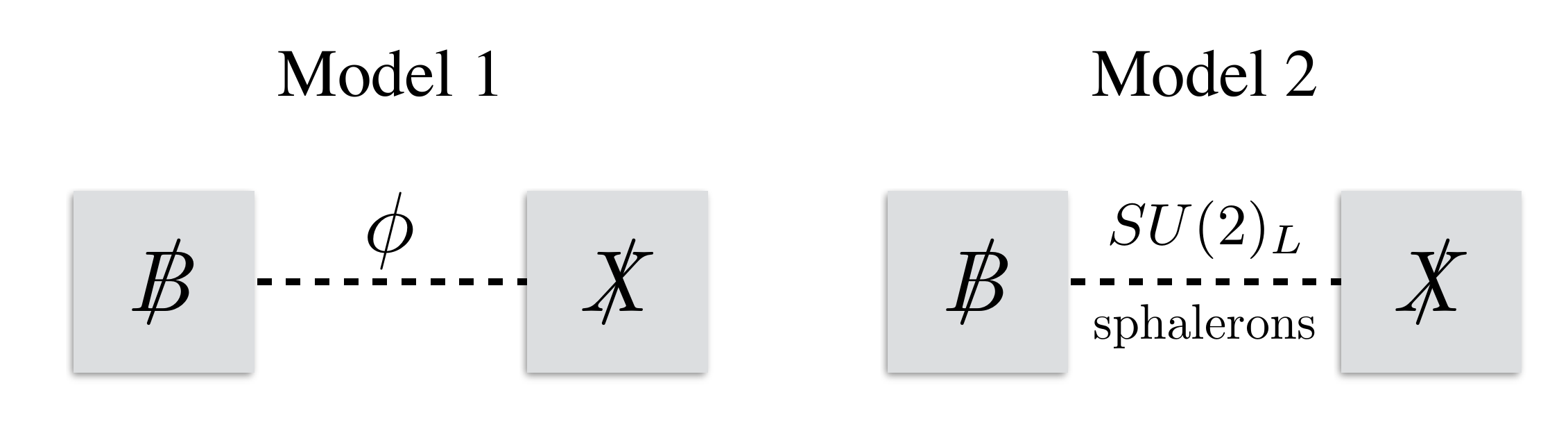}
\end{center}
\vspace*{-0.5cm}
\caption{{The two classes of ADM models considered that $\phi$ may participate in. If $\phi$ is an $SU(2)$ singlet, then $X$ and $B$ or $L$ violating operators involving $\phi$ can establish a relation between the dark and baryonic asymmetries. If $\phi$ and $X_{2}$ are doublets, DM can participate directly in the EW sphalerons and $\phi$ serves to deposit the $X_2$ asymmetry into a SM singlet fermion $X_{1}$, via $X_{2} \rightarrow \phi^{(*)} + X_{1}$. In this latter case, the additional fermion $X_1$ is needed since $SU(2)$ doublet DM is ruled out by direct detection for a wide range of masses.} } 
\label{fig:sum}
\end{figure} 
 
In this paper we 
explore the possible connection between the asymmetric origin of DM and the excess of diphotons at LHC, interpreted as a new scalar resonance. The paper is organized as follows: In Section~\ref{sec:Interpretation of the diphoton excess} we provide a brief summary of the diphoton excess interpreted in terms of a spin-0 state produced via gluon or photon fusion. In Section~\ref{sec:Implications of the Resonance for Annihilation and Asymmetries} we summarize two scenarios for generating dark matter from an initial baryon asymmetry as well as possible implications of the resonance for the annihilation of DM in the early universe.  In Section~\ref{sec:singlet} we discuss singlet models of asymmetry transfer, while in Section~\ref{sec:doublet} we discuss doublet models that use $SU(2)$ sphalerons to connect the dark and visible asymmetries and show that the diphoton resonance can be accommodated in such models. Finally we summarize our results in Section~\ref{sec:doublet}. 

\section{Interpretation of the diphoton excess}
\label{sec:Interpretation of the diphoton excess}
The diphoton excess reported by the ATLAS and CMS collaborations can be interpreted in a variety of ways~\cite{Franceschini:2015kwy,Knapen:2015dap,Backovic:2015fnp,Ellis:2015oso,Bellazzini:2015nxw,Gupta:2015zzs,Higaki:2015jag,McDermott:2015sck,Low:2015qep,Petersson:2015mkr,Cao:2015pto,Dutta:2015wqh,Kobakhidze:2015ldh,Cox:2015ckc,Ahmed:2015uqt,Becirevic:2015fmu,No:2015bsn,Demidov:2015zqn,Chao:2015ttq,Fichet:2015vvy,Curtin:2015jcv,Bian:2015kjt,Chakrabortty:2015hff,Csaki:2015vek,Alves:2015jgx,Antipin:2015kgh,Bai:2015nbs,Bardhan:2015hcr,Barducci:2015gtd,Benbrik:2015fyz,Bernon:2015abk,Carpenter:2015ucu,Chang:2015bzc,Chang:2015sdy,Chao:2015nsm,Cho:2015nxy,Dhuria:2015ufo,Ding:2015rxx,Falkowski:2015swt,Feng:2015wil,Gabrielli:2015dhk,Han:2015cty,Han:2015dlp,Han:2015qqj,Kim:2015ron,Liao:2015tow,Luo:2015yio,Megias:2015ory,Wang:2015kuj,
Altmannshofer:2015xfo,Badziak:2015zez,Bauer:2015boy,Belyaev:2015hgo,Berthier:2015vbb,Bi:2015uqd,Boucenna:2015pav,Cao:2015twy,Cao:2015xjz,Chakraborty:2015gyj,Chala:2015cev,Cline:2015msi,Cvetic:2015vit,deBlas:2015hlv,Dev:2015isx,Dey:2015bur,Gu:2015lxj,Heckman:2015kqk,Hernandez:2015ywg,Huang:2015evq,Huang:2015rkj,Kim:2015ksf,Murphy:2015kag,Patel:2015ulo,Pelaggi:2015knk,Allanach:2015ixl,Anchordoqui:2015jxc,Cai:2015hzc,Cao:2015apa,Cao:2015scs,Casas:2015blx,Chao:2015nac,Cheung:2015cug,Craig:2015lra,Das:2015enc,Davoudiasl:2015cuo,Dev:2015vjd,Gao:2015igz,Goertz:2015nkp,Hall:2015xds,Kim:2015xyn,Li:2015jwd,Liu:2015yec,Park:2015ysf,Salvio:2015jgu,Son:2015vfl,Tang:2015eko,Wang:2015omi,Zhang:2015uuo,Bizot:2015qqo,Chao:2016mtn,Chiang:2015tqz,Csaki:2016raa,Danielsson:2016nyy,Dasgupta:2015pbr,Ghorbani:2016jdq,Han:2016bus,Hernandez:2015hrt,Hernandez:2016rbi,Huang:2015svl,Ibanez:2015uok,Jung:2015etr,Kanemura:2015bli,Kaneta:2015qpf,Kang:2015roj,Karozas:2016hcp,Ko:2016lai,Low:2015qho,Modak:2016ung,Nomura:2016fzs,Palti:2016kew,Molinaro:2016oix,Kats:2016kuz,Perelstein:2016cxy,Bardhan:2016rsb,Sanz:2016auj,Panico:2016ary}.

Here we assume the diphoton excess is due to a new (pseudo) scalar state $\phi$ produced via gluon (and photon) fusion. The diphoton coupling is enhanced via the presence of new weakly charged states near the weak scale while we do not assume the presence of new colored states at the weak scale. 

Below electroweak symmetry breaking the Yukawa interactions of $\phi$ with fermions can be summarized as
\begin{align}
 {\cal L}_{\phi}  = \sum_f y_{\phi f} \, \phi \, \bar{f} \, \Gamma \, f 
\end{align}
where $f$ denotes any SM or new fermions, and
$\Gamma\equiv \{i \gamma^5, 1\} $ in the case of a (pseudo) - scalar resonance. The SM Higgs corresponds to $\Gamma=1, y_{\phi f}=m_f/v_{\rm EW}\simeq 1/\sqrt{2}$ for the SM fermions. 
The Yukawa interactions give rise to the partial widths  
\bea
\Gamma_{\phi \to \bar{f}f} &\simeq&   \frac{N_c(f) }{8\pi} y_{\phi f}^2\,  m_\phi 
\\
\label{Eq:gammapartial}
\Gamma_{\phi \gamma\gamma} &=&
\frac{\alpha^2 G_F}{128 \sqrt{2}\pi^3} m_\phi^3
\left| \sum_f N_c(f) e_f^2 c_{\phi f} A^\phi_f(\tau_f)
\right|^2
\eea
where $N_c(f) $ denotes the fermions and $c_{\phi f}$ are reduced Yukawa couplings normalized to $\frac{m_f}{v_{\rm EW}}$ following the notation in ~\cite{Djouadi:2005gj}. The loop function for a $CP$ odd $\phi$ is
\begin{align}
A^\phi_f(\tau_f) = 2 \tau^{-1} f(\tau) , \quad \tau_f=\frac{m^2_\phi}{4 m_f^2} Ê, \quad  f(\tau) =
    \begin{cases}
      \text{arcsin}^2(\sqrt{\tau}), &  \tau\leq 1 \\
      -\frac{1}{4}\left[\log\left(\frac{1+\sqrt{1-\tau^{-1}}}{1-\sqrt{1-\tau^{-1}}} \right)-i\pi \right]^2, &  \tau>1.
    \end{cases}
\end{align}

The LHC production cross section of $\phi$ via top-induced gluon fusion and/or photon fusion may be approximated by 
\bea
\label{Eq:prod}
\sigma_{\gamma\gamma}&\simeq &(\sigma_{gg\to \phi, 0} \, y_{\phi t}^2 +
\sigma_{\gamma\gamma\to \phi , 0}
\frac{\Gamma_{\phi \to \gamma\gamma}}{\Gamma_{\gamma\gamma, 0 }} ) 
\, {\rm Br}_{\phi  \to \gamma\gamma}\ . 
\eea
From \cite{Franzosi:2016wtl} we have that the gluon-fusion reference cross-section is $\sigma_{gg\to \phi, 0}=1.9$ pb for a pseudo-scalar with $y_{\phi t}=1\simeq \sqrt{2} \, y_{h_{\rm SM} t}$ where $h_{SM}$ is the the SM Higgs. The photon-fusion reference cross-section is $\sigma_{\gamma\gamma\to \phi , 0}\simeq 3$ fb for a pseudo-scalar with  $\Gamma_{\gamma\gamma , 0}=0.34$ GeV \cite{Franzosi:2016wtl} but the error on this estimate is large and we refer the reader to recent studies \cite{Fichet:2015vvy,Csaki:2015vek,Sahin:2016lda,Fichet:2016pvq,Harland-Lang:2016apc,Molinaro:2015cwg}. .

To address the observed signal cross-section $\sigma_{\gamma\gamma}\sim 5-10$ fb via the gluon fusion contribution we need $y_{\phi t}^2 \gtrsim 2 \times 10^{-3}$. At this minimum possible value for the gluon fusion production cross-section we require ${\rm Br}_{\phi\to \gamma\gamma} \simeq 1$ and thus  $\Gamma_{\phi \to \gamma\gamma} > \Gamma_{\phi \to tt}\simeq 2.5 \times10^{-4} m_\phi \sim 0.2$ GeV (for $y_{\phi t}^2 \simeq 2 \times 10^{-3}$). However, it follows Eq.~(\ref{Eq:prod}) that at this minimum value of the $y_{\phi t}$ coupling, the photon fusion can contribute comparably, see e.g.  \cite{Franzosi:2016wtl} for details.

From the diphoton decay rate in Eq.~(\ref{Eq:gammapartial}) we have that $\Gamma_{\phi \gamma\gamma} \sim 10^{-7} m_\phi \left| \sum_f N_c(f) e_f^2 c_{\phi f} A^\phi_f(\tau_f)
\right|^2$ for $m_\phi\simeq 750$ GeV and thus we need  
\begin{align}
\left| \sum_f N_c(f) e_f^2 c_{\phi f} A^\phi_f(\tau_f) \right|  \,  \gtrsim  \, 30
\end{align}
which is possible to achieve given the upper limit of $|A^\phi_{f}(\tau)|\lesssim 5$ at threshold and $y_{\phi f}\lesssim 4\pi$ from perturbativity. 
We display sufficiently large diphoton partial widths in an explicit model in Fig.~\ref{fig:Gamma} and give the production cross-section in Eq~(\ref{Eq:diphotonscaling}).

\section{Implications of the Resonance for Annihilation and Asymmetries}
\label{sec:Implications of the Resonance for Annihilation and Asymmetries}

The observation that the cosmological abundance of DM and baryons are similar
\be 
\Omega_{DM} \simeq 5~ \Omega_{B},
\ee
may imply that they shared a common origin. Since we know that the baryon abundance is related to an asymmetry, a natural possibility is to consider models which relate the baryon asymmetry to a DM asymmetry (for reviews see~\cite{Petraki:2013wwa,Zurek:2013wia}). This requires that DM is a complex  
particle carrying a particle number, $X$-number, e.g. a (pseudo-) Dirac particle charged under a global $U(1)_X$ symmetry. The key ingredient in such models are transfer operators of the form
\be
\mathcal{O}_{{\rm tr}} = \mathcal{O}_{B-L} \mathcal{O}_{X} \ , 
\label{eq:ADM}
\ee
where $\mathcal{O}_{B-L}$ is a $(B-L)$ carrying operator and $\mathcal{O}_{X}$ carries nonzero DM number. Thus $\mathcal{O}_{{\rm tr}}$  relates an asymmetry amongst baryons $\eta_{B}$ to a DM asymmetry $\eta_{X}$ by establishing chemical equilibrium between the two sectors. These operators can be the electroweak sphalerons or other higher-dimensional operators that freeze-out at relatively high scales. Often the scales are so high that DM is still relativistic, and consequently the baryon and DM number densities are similar, $n_{X} \sim n_{B}$. In this case the cosmological observation that $\Omega_{DM} /\Omega_{B} \sim 5$ implies $m_{X}  \sim \mathcal{O}(5~{\rm GeV})$ in the absence of a symmetric component of DM or multiple components. We give an explicit model of the diphoton excess leading to this relation in section~\ref{sec:doublet}.
However, importantly, larger DM masses can also be consistent with the desire to achieve $\Omega_{DM} /\Omega_{B} \sim 5$, in models where the asymmetry transfer decouples when DM has already started to become non-relativistic~(see e.g.~\cite{Barr:1990ca}). In what follows we will keep the DM mass free in order to be as general as possible.

Having now established a relation between $\eta_{B}$ and $\eta_{X}$, the final ingredient is to relate the particle asymmetries to the mass density ratio of DM and baryons~\cite{Graesser:2011wi}
\be 
\frac{m_{X}}{m_{p}} \frac{\eta_{X}}{\eta_{B}} = \left(\frac{1-r_{\infty}}{1+r_{\infty}}\right)\frac{\Omega_{DM}}{\Omega_{B}}
\label{eq:abundance}
\ee
where $r_{\infty} \equiv n_{-}/n_{+}$ with $n_{\pm}$ being the number density of (anti-)DM. Notice that the fractional asymmetry, $r_{\infty}$ is not uniquely determined by the DM asymmetry $\eta_{X}$ but instead also depends on the annihilation cross section~\cite{Griest:1986yu,Belyaev:2010kp,Graesser:2011wi}
\be
r_{\infty} \simeq \exp \left(\frac{- 0.264~\eta_{X} ~M_{{\rm Pl}}m_{X} \sigma_{0} \sqrt{g_{*}}}{x_{f}^{n+1}\left(n+1 \right)}\right),
\label{eq:r}
\ee
where $g_{*}$ is the relativistic degrees of freedom, $M_{{\rm Pl}}$ is the Planck mass, and $\langle \sigma v\rangle = \sigma_{0}(T/m_{X})^{n}$ where $n=0$ and $n=1$ are for $s$- and $p$-wave annihilation respectively. Lastly, $x_{f} = m_{X}/T_{f}$ is a dimensionless measure of the DM freeze-out temperature when DM annihilation processes cease being more rapid than the Hubble rate. This number is only logarithmically dependent on the DM mass and cross section, being typically $x_{f} \simeq 20$. 

The requirement on the annihilation cross section can be understood analytically in the small $r_{\infty}$ limit~\cite{Lin:2011gj} by combining Eq.(\ref{eq:abundance}) and (\ref{eq:r})
\be 
\langle \sigma v_{{\rm rel}} \rangle_{ADM} \simeq 5 \times 10^{-26}~{\rm cm}^{3}~{\rm s}^{-1}~\log\left(\frac{1}{r_{\infty}}\right),
\ee
such that larger annihilation cross section result in much smaller fractional asymmetries.  This is in line with the generic expectation that ADM models require larger than WIMP-sized annihilation cross sections since the ``symmetric component'' needs to be annihilated away for the asymmetric excess to account for the DM abundance. 

The preceding discussion applies to a wide class of ADM models. Now let us investigate the implications of the new resonance state for ADM.
\subsection{Generic Implications}

 \begin{figure}[t]
\begin{center}
\includegraphics[width=.45\textwidth]{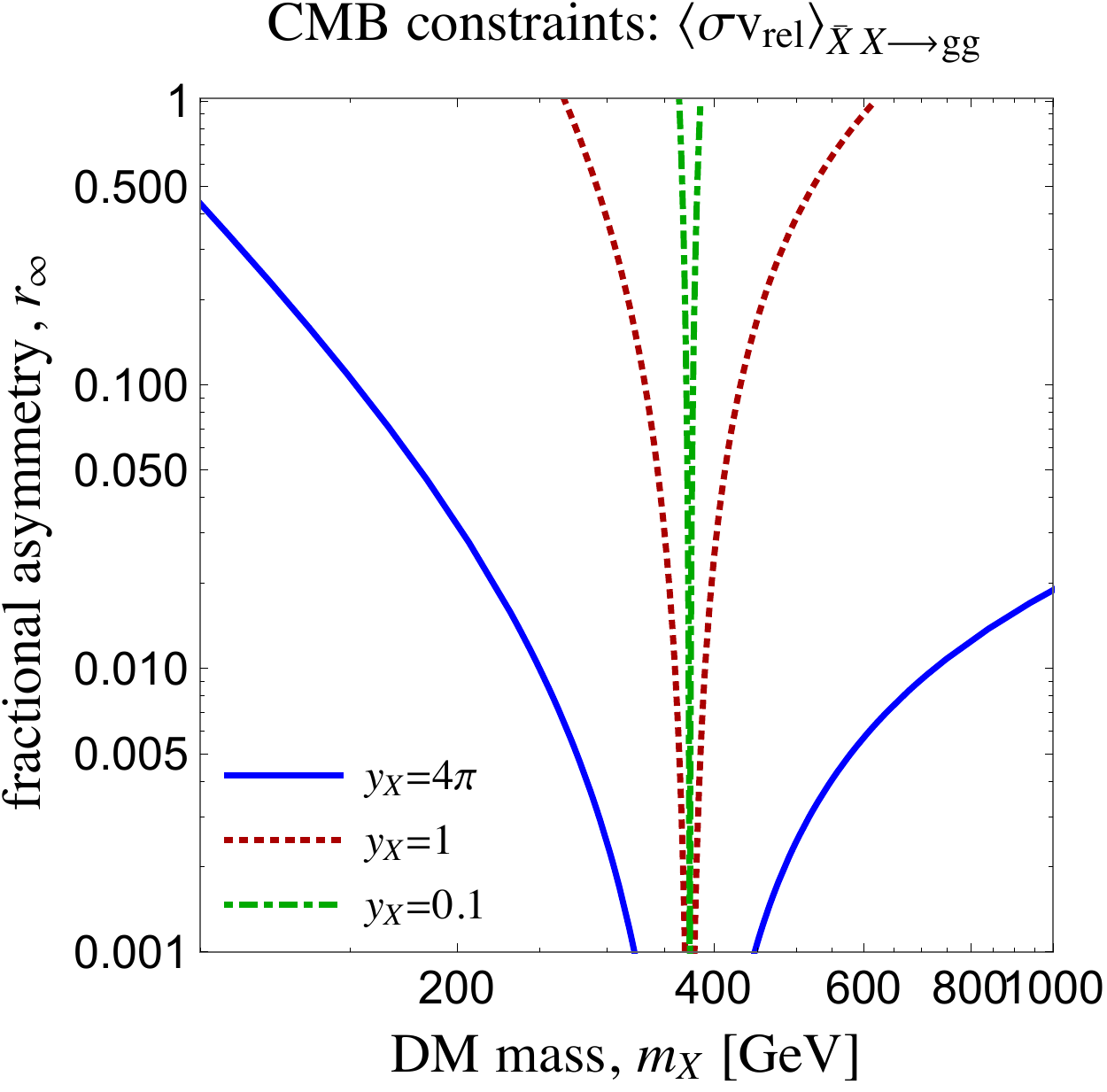}
\includegraphics[width=.45\textwidth]{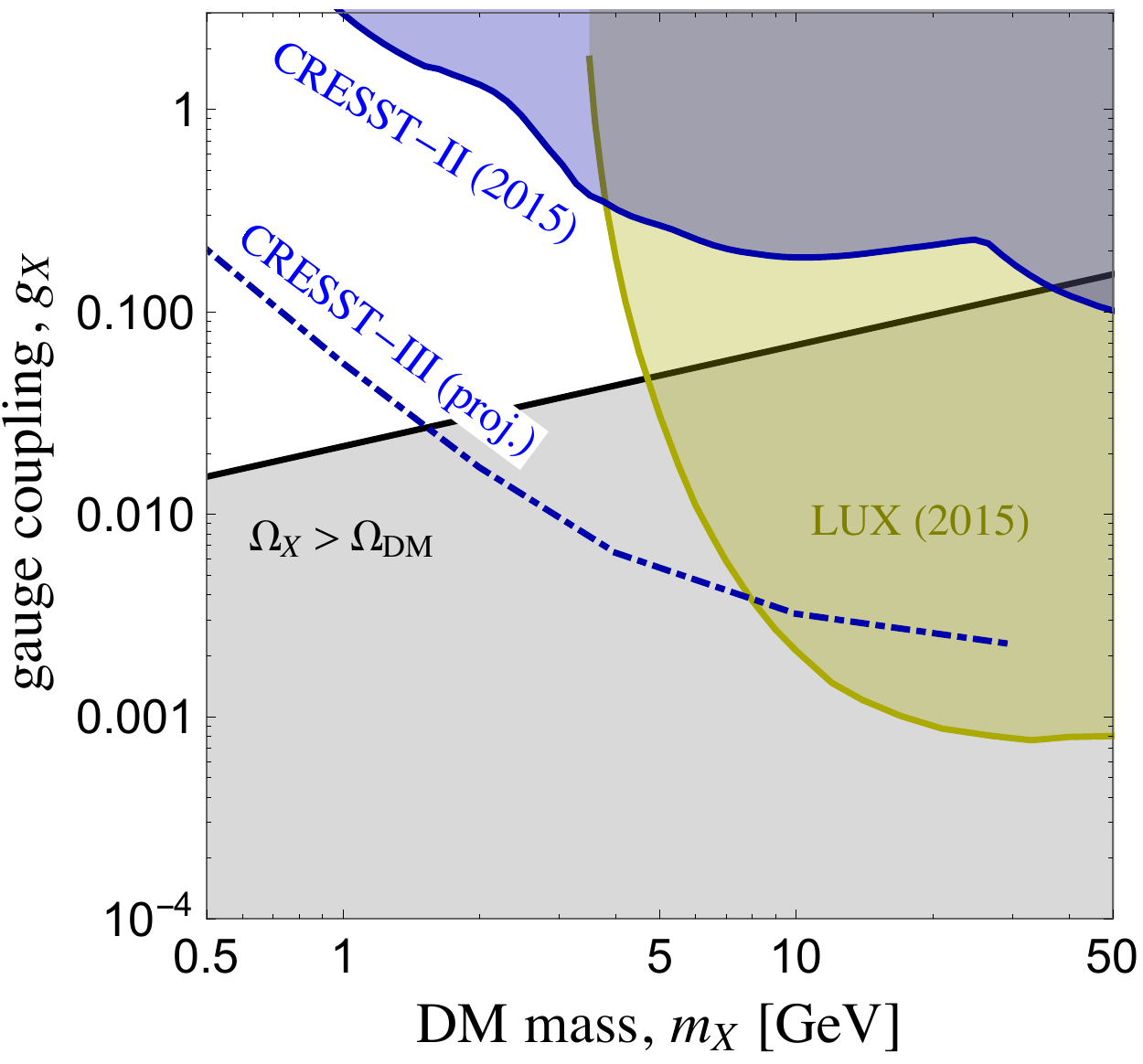}
\end{center}
\vspace*{-0.5cm}
\caption{{\it Left panel:} Annihilation to gluons in the CMB, taking the gluon-$\phi$ coupling $\frac{1}{\Lambda} \phi G_{\mu \nu} \tilde{G}^{\mu \nu}$ to be $\Lambda = 5$ TeV~\cite{D'Eramo:2016mgv}. {\it Right panel:} Current and future direct detection limits on a gauged $U(1)_{X}$ model of the DM relic abundance. In addition to current LUX~\cite{Akerib:2015rjg} and CRESST-II~\cite{Angloher:2015ewa} limits we also display a CRESST-III phase 2 projection based on 100 eV threshold with 1000 kg-day exposure~\cite{cresst3}. Here we have fixed the kinetic mixing parameter $\varepsilon = 10^{-9}$ and the vector mass $m_{A'} = 10$ MeV. }
\label{fig:annihilation}
\end{figure}

As discussed above, ADM models must feature a sizeable DM annihilation cross section in order to remove the symmetric component~\cite{Belyaev:2010kp,Graesser:2011wi}. Since this condition applies 
generically we investigate it first. A simple connection of DM to the diphoton excess might be a direct coupling of DM to the new resonance. This immediately implies new annihilation channels for DM such as
\bea \bar{X} X &\rightarrow&  \phi \rightarrow gg \\
\bar{X} X &\rightarrow&  \phi \rightarrow \gamma \gamma
\eea
In addition, if DM is sufficiently heavy ($m_{X} > m_{\phi}$) then we also have $t$-channel annihilation,
\be
\bar{X} X \rightarrow  \phi \phi.
\ee
However, because of the gluon couplings these annihilation channels also imply DM signals at both hadron colliders (jet(s) plus missing energy) and elastic scattering signals at direct detection experiments. For the $\mathcal{O}(10~{\rm GeV})$ DM range, natural for ADM, these constraints rule out the above annihilation modes as the ones setting the relic density or annihilating away the symmetric component of
DM~\cite{D'Eramo:2016mgv}.  

It has been argued that late time CMB annihilation can constrain both symmetric~\cite{Padmanabhan:2005es,Galli:2011rz,Hutsi:2011vx} and asymmetric DM annihilation~\cite{Lin:2011gj,Bell:2014xta}. This constraint comes from the injection of energy into the electromagnetic plasma during recombination and therefore relies on the efficiency of energy deposition $f_{{\rm eff}}$ which depends on the DM annihilation channel. For the $\gamma \gamma$ and $gg$ final states of interest to us the efficiency is roughly independent of the DM mass and is of the size, $f_{gg} \simeq 0.2$ and  $f_{\gamma\gamma} \simeq 0.4$~\cite{Slatyer:2015jla}. 

In the ADM case, the CMB annihilation constraint can be recast as a limit on the fractional asymmetry, $r_{\infty}$,
\be 
f_{{\rm eff}} \frac{\langle \sigma v\rangle}{m_{X}} \frac{r_{\infty}}{2}\left(\frac{2}{1+r_{\infty}}\right)^{2} < 1.2 \times 10^{-27}~{\rm cm}^{3}~{\rm s}^{-1}
\ee
where the right-hand side applies the current (WMAP9+Planck+ACT+SPT+BAO+HST+SN) CMB limit at $95~\%$ CL~\cite{Madhavacheril:2013cna}. We find that gluon annihilation dominates over the photon channel in setting the most stringent CMB limit, which we display in Fig.~\ref{fig:annihilation} for a pseudo-scalar mediator which provides $s$-wave annihilation, 
\be 
\langle \sigma v_{{\rm rel}}\rangle_{gg} \simeq \frac{32 y_{X}^{2}m_{X}^{2}}{\pi \Lambda^{2}}\frac{1}{\left(4m_{X}^{2} - m_{\phi}^{2}\right)^{2} + \Gamma_{\phi}^{2}m_{\phi}^{2}}
\ee

Note that of the annihilation modes considered here those mediated by the $s$-channel pseudo-scalar are the only ones that yield strong limits from CMB data since all the others are strongly $p$-wave suppressed.

We note that all models need additional DM interactions beyond a coupling to the resonance in order to provide a viable thermal relic. This is a result of the complementarity of collider, direct and indirect astrophysical searches of DM (see e.g.~\cite{D'Eramo:2016mgv}). In particular for scalar resonances the spin-independent direct detection limits from LUX~\cite{Akerib:2015rjg} are strong enough to rule out DM masses less than $m_{\phi}/2$, while the upcoming LZ collaboration will probe the remaining window at high mass thermal DM~\cite{D'Eramo:2016mgv}. And in the case of a scalar resonance, the indirect searches are very weak since all of the minimal annihilation channels are $p$-wave suppressed. 

In contrast for pseudo-scalar resonances, indirect limits on DM annihilation can be strong while direct detection is extremely weak. The weakness of direct detection in this case is due to the momentum-suppression in the cross section as well as the fact that pseudo-scalar interactions mediate spin-dependent scattering, for which the present limits are much weaker due to the lack of coherent enhancement. As a result of the combination of LHC monojet~\cite{Aad:2015zva} and the recent Fermi $\gamma$-ray line limits~\cite{Ackermann:2015lka}, thermal DM with masses below $m_{\phi}/2$ are excluded~\cite{D'Eramo:2016mgv}. 

Thus typically one will need additional annihilation channels for the DM relic abundance.  This is easily accommodated however in models where DM is charged under a new $U(1)_{X}$ gauge symmetry with a massive gauge boson, $A_{\mu}'$ lighter than the DM mass. Interestingly this possibility endows DM with rather large velocity-dependent self-interactions, which may be suggested by the small-scale structure problems of collisions cold DM~\cite{Loeb:2010gj,Kaplinghat:2013yxa,Tulin:2013teo}. The presence of light scale vectors with sizeable kinetic mixing with the photon can be unveiled in a DM halo-independent manner with the combination of multiple detector types~\cite{Cherry:2014wia}.

{For illustrative purposes we show in the right panel of Fig.~\ref{fig:annihilation} the current and projected direct detection limits on this $U(1)_{X}$ gauge interaction model for DM annihilation. Elastic scattering proceeds via kinetic mixing with the SM photon, $\varepsilon F^{\mu \nu} F'_{\mu \nu}$ where $F'_{\mu \nu}$ is the $U(1)_{X}$ field strength. In Fig.~\ref{fig:annihilation} we fix $\varepsilon = 10^{-9}$ and $m_{A'} = 10$ MeV. We note that experiments sensitive to electron-DM scattering may offer additional constraints in the sub-GeV regime~\cite{Essig:2012yx,Hochberg:2015pha,Essig:2015cda}.}

\subsection{Model-dependent Implications for the Asymmetry}
One can establish a relation between the baryonic and DM asymmetries in one of two ways: (1) Generate $\eta_{X} \neq 0$ primordially and then transfer it to baryons via Eq.(\ref{eq:ADM}) (or vice versa); or (2) Generate both $\eta_{X} \neq0$ and $\eta_{B} \neq0$ simultaneously and circumvent the need to transfer the asymmetry.  Here we will not explore this latter possibility in much detail, but comment on the possibilities of relating the resonance with ``cogenesis'' mechanisms.   For example, in supersymmetric (SUSY) models it may be associated with a flat direction (in the SUSY preserving and renormalizable limit) of the scalar potential that allows for the generation of a large primordial asymmetry via the Affleck-Dine mechanism~\cite{Affleck:1984fy,Dine:1995kz}. Models of this type have been generalized beyond traditional baryogenesis to allow for the simultaneous production of dark and baryonic asymmetries~\cite{Bell:2011tn,Cheung:2011if,Graesser:2011vj,vonHarling:2012yn}. 

Here we shall consider two model classes that realize a connection between the diphoton excess and ADM via the transfer of the asymmetries, summarized schematically in Fig.~\ref{fig:sum} and discussed briefly below:
\begin{itemize}

\item {\bf Model 1: New Asymmetry Transfer Operators}: If the diphoton resonance and DM are both EW singlets, then EW sphalerons cannot transfer a particle asymmetry between the dark and visible sectors, but $\phi$ can play this role.
For example, two operators involving $\phi$ that accomplish this task for singlet $\phi$ and $X$ are
\bea 
\mathcal{O}_{SM} &=& \frac{1}{\Lambda_{{\rm tr}}} \phi (LH)^{2},\label{eq:transfer1}\\ 
\mathcal{O}_{X} &=&y_{X} \phi XX,
\label{eq:transfer2}
\eea
which act to establish chemical equilibrium between dark and leptonic asymmetries. As long as these interactions are in equilibrium above the EW scale, sphalerons will establish a relation between the $X$ and baryon numbers.  Note that in the limit where $\phi$ is sufficiently heavy that it can be integrated out we recover the transfer operator studied in~\cite{Cohen:2009fz}. 

We study this model in Sec.~\ref{sec:singlet}.

\item {\bf Model 2: EW Sphalerons and new $SU(2)$ Charged States}: If the diphoton resonance is part of an $SU(2)$ doublet it may signal additional EW structure. 
If DM carries weak quantum numbers and a particle number $X$ that is classically conserved, but violated by the weak anomaly, then DM can be produced asymmetrically in the EW sphaleron transitions. 
\bea 
\mathcal{O}_{Sphaleron} =  (QQQ L)^3 X_2^2 
\eea
However, as is well-known, the simplest incarnation of this setup where DM itself is part of an elementary $SU(2)$ doublet is in tension with direct detection constraints since the DM fermion participating in sphalerons also scatters on nuclei via the $Z$ boson.

A simple way to avoid this problem is to consider models in which a heavy doublet $X_{2}$ carrying $X$ number decays into the resonance scalar $\phi$ and a light singlet DM state $X_{1}$ in the early universe. In this case $X_2$ can acquire an asymmetry via sphalerons while the decay $X_{2} \rightarrow \phi X_{1}$ via
\bea 
\mathcal{O}_{{\rm decay}} =  X_2 X_1 \phi 
\eea
transfers the asymmetry into the $SU(2)$ neutral state $X_1$. The $X_{1,2},\phi$ states may be fundamental \cite{Barr:1991qn,Perez:2013tea} or composite \cite{Frandsen:2011kt}. The $\phi$ state may be identified with the SM Higgs as in \cite{Perez:2013tea} but that is now constrained by e.g. the Higgs diphoton decay rate and it is therefore relevant to identify $\phi$ with a new spin-0 doublet in addition to the SM one, as in the 2HDM. 
We study such a model in Sec.~\ref{sec:doublet}.

\end{itemize}

\section{Models of Asymmetry Transfer With Spin-0 Singlets}
\label{sec:singlet}

We follow~\cite{Harvey:1990qw} for the notation and setup of chemical equilibrium in the early universe. This particular transfer operator evolution is similar to a model considered in~\cite{Graesser:2011wi}, whose approach we also follow.  We solve for the ratio of baryon number to DM number by breaking the problem into two steps. In the first step, we find the value of the baryon density and $L'$ density ($L' = X-L$)  at the sphaleron temperature since these two quantum numbers are conserved.

\begin{figure}
\includegraphics[width=0.45\textwidth]{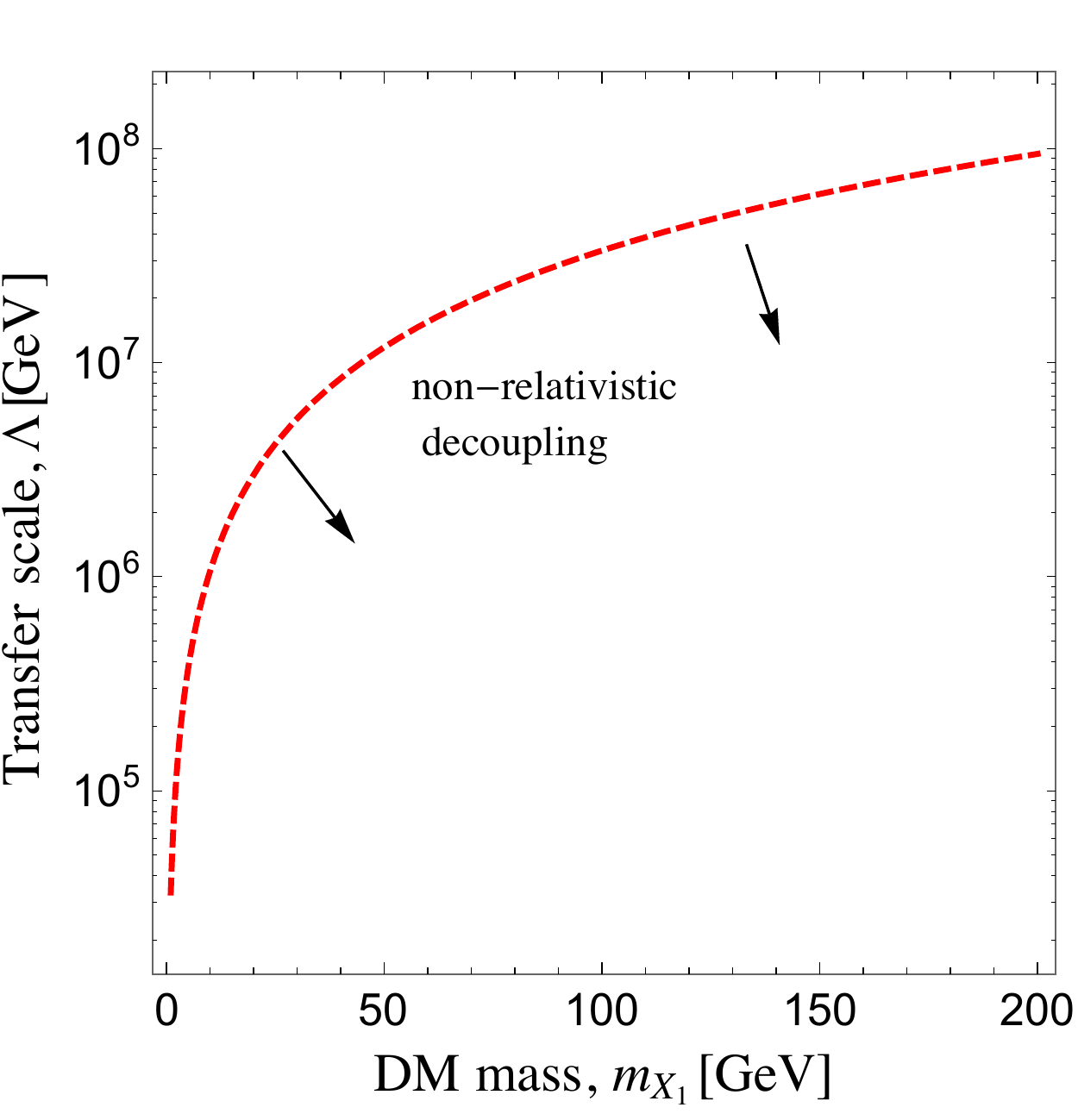}
\caption{Here we show what values of the transfer operator scale are needed in order to obtain non-relativistic decoupling for a range of DM masses. For values of the transfer scale above this critical value only rather light DM can be accommodated (see Eq.(\ref{eq:DMmass})).}
\label{fig:DMmassmodel1}
\end{figure}

The chemical equilibrium conditions are
\bea
Q &\propto& 12\mu_{u} - 6\mu_{d} - 6 \mu_{e} = 0 \\
\mu_{u} - \mu_{d} &=& \mu_{\nu} -\mu_{e} \\
\mu_{u} + 2\mu_{d} +\mu_{\nu}  &=& 0 \\
2\mu_{X} + 2\mu_{\nu}  &=& 0\\
\eea
following from EM charge conservation, $W^{\pm}$ exchange, and the transfer operator (Eqs.(\ref{eq:transfer1} and (\ref{eq:transfer2}))).  At the sphaleron decoupling scale $T_{{\rm sph}}$ these yield 
\bea 
B &=& -\frac{36}{7}\mu_{e}, \\
L' &=& \left(\frac{11}{36} \frac{f(m_{X}/T_{{\rm sph}})}{f(0)} + \frac{3}{2}\right) B,
\eea
where the function $f(x)$ encodes the Boltzmann suppression in the DM density as it becomes non-relativistic at the epoch of decoupling, 
\be 
f(x) = \frac{1}{4\pi^{2}}\int_{0}^{\infty} dy~\frac{y^{2}}{{\rm cosh}^{2}~\left(\frac{1}{2}\sqrt{x^{2}+y^{2}}\right)}.
\ee
Note that in case the of scalar DM the above function is the same modulo the replacement $\cosh(x) \rightarrow \sinh(x)$.

Next, we impose these value of $B$ and $L'$ as initial conditions for the evolution from $T_{{\rm sph}}$ through the transfer operator decoupling at $T_{D}$. Solving finally for the ratio of dark and baryon numbers we find 
\be 
\frac{\eta_{X}}{\eta_{B}} = \left( \frac{54 + 11 \frac{f(m_{X}/T_{sph})}{f(0)}}{360 + 36 \frac{f(m_{X}/T_{D})}{f(0)} } \right) \frac{f(m_{X}/T_{D})}{f(0)}
\label{eq:asymmetry1} 
\ee
Thus for example when DM is very light compared to both the sphaleron and the transfer temperatures, the DM to baryon asymmetry ratio is $\eta_{X}/\eta_{B} \simeq 0.16$ implying from Eq.~(\ref{eq:abundance}) that the DM mass can range from
\be
m_{X} = \begin{cases} 
      30.5, & r_{\infty} =0 \\
      1.6, & r_{\infty} =0.9 
   \end{cases}
   \label{eq:DMmass}
   \ee
Heavier DM can easily be arrange by allowing for the Boltzmann suppression functions in Eq.~(\ref{eq:asymmetry1}).

Lastly, for completeness we show that it is quite straightforward to obtain high-scale decoupling of the asymmetry transfer.  First, we solve for the transfer operator decoupling temperature by computing $n_{X} \sigma_{{\rm  tr}} \simeq H(T_{D})$ where the transfer operator cross section induced from Eqs.~(\ref{eq:transfer1}) and (\ref{eq:transfer2}) is
\be 
\sigma_{{\rm tr}} \simeq \left(\frac{v}{\Lambda_{{\rm tr}}}\right)^{2} \frac{y_{X}^{2}}{16 \pi m_{\phi}^{4}}m_{X}^{2}
\ee
This results in the transfer operator decoupling temperature 
\be 
T_{D} = \frac{m_{X}}{\log\left[ \left(\frac{v}{\Lambda_{{\rm tr}}}\right)^{2} \frac{y_{X}^{2}M_{Pl}}{16 \pi m_{\phi}^{4}}m_{X}\right]}.
\ee
We thus see that DM masses in the 1 GeV - 1 TeV range are well accommodated for even very high scales of the transfer operator $\Lambda_{{\rm tr}}$. This makes the transfer component of the model quite safe from experimental scrutiny, though we note that future direct detection constraints can limit the Yukawa $y_{X}$ involved directly in the operator $\mathcal{O}_{X}$, in Eq.~(\ref{eq:transfer2}). Moreover, here we have assumed that $\phi$ does not obtain a vacuum expectation value. If that assumption is broken however then Eq.~(\ref{eq:transfer2}) induces a Majorana mass for the DM allowing for $X-\bar{X}$ oscillations~\cite{Cohen:2009fz,Buckley:2011ye,Cirelli:2011ac,Tulin:2012re}. The final relic abundance can be impacted in this case depending on the precise scale of the Majorana term, and of course present-day annihilation can proceed at sizeable rates.

Finally of course the available annihilation channels in the minimal setup for the transfer operator and diphoton signal and not sufficient for the DM relic abundance. As mentioned earlier however, this is achieved in models where DM is charged under a new gauge symmetry that has low-scale gauge boson, $m_{A'} \ll m_{X}$. Data from direct detection experiments with disparate target media may be able to constrain the mass of this gauge boson via the momentum dependence in the elastic scattering cross section~\cite{Cherry:2014wia}.

\section{Models of Asymmetry Transfer With Spin-0 Doublets}
\label{sec:doublet}

We now discuss models of ADM such as those in~\cite{Barr:1991qn,Frandsen:2011kt,Perez:2013tea} featuring new spin-0 and spin-1/2 weak doublets allowing to interpret the diphoton excess as a pseudo-scalar resonance.  As mentioned earlier, in this case an $X$-charged EW doublet (fermion) state can participate in the sphaleron transitions and generate a nonzero DM number in tandem with a nonzero baryon number.  To evade direct detection constraints though, this state cannot be the DM today. Instead the heavier $X$-carrying state decays to EW singlet DM.  We assume the production of a baryon asymmetry at some high scale and only require the model be able to transfer this asymmetry. 

As a concrete example consider introducing the following left handed fermions, 
\begin{gather}
\begin{aligned}
\Psi_L=\begin{pmatrix} \psi_L^0 \\ \psi_L^- \end{pmatrix}   &\sim  ( 2, -\frac{1}{2}, 1),\\
\eta_L^+   &\sim  ( 1, 1, -1),\\
\eta_L   &\sim  ( 1, 0, -1),\\
\widetilde{\chi}_L &\sim ( 1, 0, -1), \hspace{3mm} \text{and}
\end{aligned}\hspace{3mm}
\begin{aligned}
\widetilde{\Psi}_L=\begin{pmatrix} \widetilde{\psi}_L^+ \\ \widetilde{\psi}_L^0 \end{pmatrix} &\sim (\bar{2}, \frac{1}{2}, 1), \\
\widetilde{\eta}_L^- &\sim  (1, -1, -1), \\ 
\widetilde{\eta}_L &\sim  (1, 0, -1), \\ 
\chi_L &\sim  ( 1, 0, 1)\,,
\label{Eq:doubletmodel}
\end{aligned}
\end{gather}
charged under the weak gauge symmetries and a global $U(1)_X$ symmetry,
 \begin{equation*}
SU(2)_L \otimes U(1)_Y \otimes [U(1)_X] .
 \end{equation*}
The second doublet $\widetilde{\Psi}_L$ is here needed for anomaly cancellation, and we introduced the right handed fields $\eta,\widetilde{\eta}$ to allow Dirac masses for both the charged and neutral components. 

We also introduce two scalar doublets $\Phi_i, i=1,2$ and a scalar singlet $S$
\begin{eqnarray}
\Phi_i \sim  ( 2, -\frac{1}{2},0) = \begin{pmatrix} \phi_i^0 \\ \phi_i^- \end{pmatrix} , 
\qquad  S \sim(1,0,0)  Ê.
\end{eqnarray}
The doublet $\Phi_2$ is coupled to the SM fermions, and yields the (mostly) SM higgs doublet, while $\Phi_1$ is coupled only to the new weak doublets. The scalar sector is thus an extension of a Type-I two Higgs doublet model. We also introduce Yukawa interactions of $\Phi_1$ with the new fermions 
\begin{eqnarray}
-\mathcal{L} &\supset&  
\label{Eq:Yukawaphoton1}
y_1( \widetilde{\Phi}_{1}^\dagger \Psi_{L} \eta_L^+ + \Phi_{1}^\dagger  \widetilde{\Psi}_{L}  \widetilde{\eta}_L^- ) + {\rm h.c. } \\
\label{Eq:Yukawaphoton2}
& + & y_2 (\Phi_{1}^{\dagger} \Psi_{L}\eta_L  +  \widetilde{\Phi}_{1}^\dagger \widetilde{\Psi}_{L}  \widetilde{\eta}_L)  +  {\rm h.c. } \\
\label{Eq:Yukawaeq1}
&+& y_3 (\Phi_{1}^{\dagger} \Psi_{L} \widetilde{\chi}_L  +  \widetilde{\Phi}_{1}^\dagger \widetilde{\Psi}_{L}  \widetilde{\chi}_L)    +  {\rm h.c. } \\
& + & \lambda_S S  \widetilde{\chi}_L \chi_L  + {\rm h.c. }
\end{eqnarray}
Below the EW scale we then identify two electrically charged Dirac fermions, carrying also $X$-charge, as 
\begin{align}
X_2^-=\begin{pmatrix} \psi_L^- \\ \eta_L^{+ \dagger} \end{pmatrix}  , \quad 
\widetilde{X}_2^-=\begin{pmatrix} \widetilde{\psi}_L^+ \\ \widetilde{\eta}_L^{-\dagger} \end{pmatrix} 
\end{align}
These will contribute to the diphoton decay of the (dominantly) $\Phi_1$ physical scalars. 

Taking for simplicity $y_1=y_2=y$ and $y_3,\lambda_s$ to be parametrically smaller than $y$ the heavy charged and neutral (approximate) mass eigenstates have the common mass
\begin{align}
m_{X_2} =m_{\widetilde{X}_2}=\frac{y v_1}{\sqrt{2}}
\end{align}
where the heavy neutral mass eigenstates to zeroth order in $y_3/y, \lambda_s/y$ are 
\begin{align}
X_2^0\simeq \begin{pmatrix} \psi_L^0 \\ \eta_L^{\dagger }  \end{pmatrix}  , \quad 
\widetilde{X}_2^0\simeq \begin{pmatrix} \widetilde{\psi}_L^0 \\ \widetilde{\eta}_L^{\dagger }  \end{pmatrix}  \ , 
\end{align}
while the light neutral mass eigenstate is
\begin{align}
X_1 \simeq \begin{pmatrix} \chi_L \\ \widetilde{\chi}_L^\dagger \end{pmatrix}  , \quad m_{X_1} \simeq \lambda_s v_s   
\end{align}

Electroweak sphalerons and the associated `t Hooft operator 
\begin{equation}
(u_L d_L d_L \nu_L)^3\overline\Psi_R\Psi_L = (u_L d_L d_L \nu_L)^3\widetilde{\Psi}_L\Psi_L   \,.
\end{equation}
violate $X$-number by two units. The origin of the DM particle $\widetilde{\chi}_L$ is therefore as follows. 

i) In the early universe we assume an initial baryon asymmetry is produced. ii) Below this scale $B-L$ is conserved, but sphalerons transfer the initial asymmetry into $\Psi_L,\widetilde{\Psi}_L$. iii) Below the sphaleron freeze out the $\Psi_L,\widetilde{\Psi}_L$ states decay into the mass eigenstate $X_{1R}^\dagger \simeq \widetilde{\chi}_L$  via Eq.~(\ref{Eq:Yukawaeq1}).

We note that the specific Yukawa structure we assume above, notably the distinction between the neutral $\eta$ and $\chi$ states can be ensured by endowing the new fermions, except for $\chi, \widetilde{chi}$ with lepton number as done in \cite{Barr:1991qn}, at the expense of having to introduce another lepton charged scalar doublet mediating the interaction in Eq.~(\ref{Eq:Yukawaeq1}). 
Another relevant variation of the model is given in \cite{Perez:2013tea} where baryon number instead of $X$ number is assigned to the states.

The ratio of DM number $X$ and baryon number $B$ in the model above is
\begin{align}
\frac{X}{B} \simeq (1 + 4 f(m_{X_2}/T^*)/f(0))
\end{align}

such that using Eq.~(\ref{eq:abundance}) we arrive at 
\be
m_{X_1}
\simeq \begin{cases} 
      5 & r_{\infty} =0 \\
      0.25, & r_{\infty} =0.9. 
   \end{cases}
   \label{eq:DMmass2}
   \ee
Thus in contrast with the singlet model of Sec.~\ref{sec:singlet}, we achieve much lighter DM in the case of relativistic transfer decoupling and the initial $(B-L)$ charge vanishes.  The variations in \cite{Barr:1991qn,Perez:2013tea} produce a similar mass for the DM candidate while \cite{Frandsen:2011kt} provides a composite realization.

\subsection{Scalar sector}
\label{Scalar sector}
We now investigate the scalar sector of the model with the aim of explaining the LHC diphoton excess. The scalar sector is a 2 Higgs doublet model (2HDM)--- augmented by a singlet scalar $S$ and new weak charged fermions. 
We first disregard effects of the scalar $S$ assuming it is only weakly coupled to the 2HDM and consider the $CP$-conserving potential,  e.g. \cite{Gunion:2002zf}.
\begin{eqnarray}
\label{Eq:scalarpotential}
\mathcal{V}&&= m_{11}^2\Phi_1^\dagger\Phi_1+m_{22}^2\Phi_2^\dagger\Phi_2
-[m_{12}^2\Phi_1^\dagger\Phi_2+{\rm h.c.}]\nonumber\\[8pt]
&&+\frac{1}{2}\lambda_1(\Phi_1^\dagger\Phi_1)^2
+\frac{1}{2}\lambda_2(\Phi_2^\dagger\Phi_2)^2
+\lambda_3(\Phi_1^\dagger\Phi_1)(\Phi_2^\dagger\Phi_2)
+\lambda_4(\Phi_1^\dagger\Phi_2)(\Phi_2^\dagger\Phi_1)
\nonumber\\[8pt]
&&+\left\{\frac{1}{2}\lambda_5(\Phi_1^\dagger\Phi_2)^2
+\big[\lambda_6(\Phi_1^\dagger\Phi_1)
+\lambda_7(\Phi_2^\dagger\Phi_2)\big]
\Phi_1^\dagger\Phi_2+{\rm h.c.}\right\}\,. \label{pot}
\end{eqnarray}

We define two angles 
$\beta$, the ratio of the scalar vevs
\begin{equation}
\tb\equiv\tanb\equiv{v_2\over
  v_1}\, ,
\label{tanbdef}
\end{equation}
with $0\leq \beta \leq \pi/2$. And $\alpha$  determining the `Higgs mixtures' upon diagonalization. The physical states --- the $CP$-even higgs scalar $h$ and heavy scalar $H$, the $CP$-odd $A$ and the charged scalars $H^{\pm}$ --- as well as the absorbed Goldstone bosons $G$ are then given by 
\begin{eqnarray}
\Phi_1^\pm &=& \cb G^\pm-\sb H^\pm\,,\nonumber \\
\Phi_2^\pm &=& \sb G^\pm+\cb H^\pm\,,\nonumber \\
\Phi_1^0 &=& \nicefrac{1}{\sqrt{2}}\left[v_1+\ca H-\sa h +i\cb
G-i\sb\ha\right]\,,\nonumber \\ 
\Phi_2^0 &=& \nicefrac{1}{\sqrt{2}}\left[v_2+\sa H+\ca h +i\sb
G+i\cb\ha\right]\,.
\label{heigenstates}
\end{eqnarray}

For $\alpha \to 0, \beta \to \pi/2$ we have that $\Phi_2$ is a pure SM Higgs. We may approach this decoupling limit via the small parameter $\delta\equiv\beta-\alpha-\pi/2$, with $\cba \simeq -\delta \ll 1, \sba \simeq 1$ to first order.

\subsubsection{Spectrum}
In the decoupling limit $\delta \ll 1$ it is convenient to express the spectrum in terms of the following parameter combinations \cite{Gunion:2002zf}:
\begin{eqnarray}
\mha^2 &\simeq &
v^2 \left[{\lamhat\over\cbma}+\lam_A-\nicefrac{3}{2}\lamhat\,\cbma
\right]\simeq  v^2 \left[- {\lamhat\over\delta}+\lam_A+\nicefrac{3}{2}\lamhat\,\delta
\right]
\,,\label{decouplimits1}\\
m_h^2 &\simeq & v^2(\lambar-\lamhat\,\cbma) \simeq  v^2(\lambar + \lamhat\,\delta) 
\,,\label{decouplimits2}\\
m_H^2 &\simeq &
v^2 \left[{\lamhat\over\cbma}+\lam-\frac{1}{2}\lamhat\,\cbma
\right]\simeq \mha^2
+(\lambda-\lambda_A - \lamhat\,\delta)v^2\,,\label{decouplimits3}\\
m_{H^\pm}^2 &\simeq &
v^2 \left[{\lamhat\over\cbma}+\lam_A+\frac{1}{2}\lambda_F
-\nicefrac{3}{2}\lamhat\,\cbma\right]=\mha^2+\frac{1}{2}\lambda_F v^2\,.
\label{decouplimits4}
\end{eqnarray}
where
\begin{eqnarray}
\!\!\!\!\!\!\!\!\!\!\!\!\lambar&\equiv& \lam_1\cb^4+\lam_2\sb^4
+\frac{1}{2}\lambda_{345}\stwob^2
+2\stwob(\lam_6\cb^2+\lam_7\sb^2)
\label{lambardef}
\,,   \\[5pt]
\!\!\!\!\!\!\!\!\!\!\!\!\lamhat&\equiv& \frac{1}{2}\stwob\left[\lam_1\cb^2
-\lam_2\sb^2-\lambda_{345}
  \ctwob\right]-\lam_6\cb\cthreeb-\lam_7\sb\sthreeb
\label{lamhatdef}
\,,\\[5pt]
\!\!\!\!\!\!\!\!\!\!\!\!\lam_{\ha}&\equiv&
\ctwob(\lam_1\cb^2-\lam_2\sb^2)+\lambda_{345}s_{2\beta}^2-\lam_5
+2\lam_6\cb\sthreeb-2\lam_7\sb\cthreeb\,,\label{lamadef} \\[5pt]
\!\!\!\!\!\!\!\!\!\!\!\!\lambda_F&\equiv&\lam_5-\lam_4\,,
\label{lamfdef}
\end{eqnarray}
and $\lambda_{345}\equiv \lambda_{3}+ \lambda_{4}+\lambda_{5}$.

Thus from the observation of the 125 GeV scalar and the putative 750 GeV state that we will identify with the $CP$-odd scalar $A$ we require 
\begin{eqnarray}
(\lambar+\lamhat\,\delta) \simeq 1/4 \, , \quad - {\lamhat\over\delta} \sim 10
\end{eqnarray}
Further we require that the decay mode from the $A\to ZH$ coupling is negligible, thus $m_H\geq m_A$ and for $H$ to contribute to the signal also $M_H-M_A < M_V$, thus 
\begin{eqnarray}
\frac{(\lambda-\lambda_A+\lamhat\,\cbma)v^2}{(M_H+M_A)M_Z} = 
\frac{m_h^2 + (2 \lamhat\,\cbma -\lambda_A)v^2}{(M_H+M_A)M_Z}  \lesssim 1
\end{eqnarray}

A parameter example that satisfies these requirements, and is in agreement with vacuum stability conditions, is $\delta \simeq -0.1, \hat{\lambda}\simeq1, \lambda=\lambda_A\simeq1/3$.

\subsection{Couplings and diphoton rates}
\label{Couplings and diphoton rates}
The 2HDM we consider is Type-I like with only $\Phi_2$ coupled to the SM fields. 
 The reduced couplings, normalized to the SM higgs couplings, of the 125 GeV scalar $h$ are then to first order in $\delta$, see e.g. \cite{Alves:2012ez,Mahmoudi:2009zx}, 
\begin{align}
c_{h f}&  = \ca / \sb = 1 - \cotb \, \delta
\\
c_{h V}&= \ca \sb - \sa \cb = \sba = 1 \ , 
\end{align}
which for $\cot\beta< 1$ and $\delta \sim O(10^{-1})$ are very close to the SM higgs values. 
Similarly the couplings of the heavy neutral scalar $H$ and pseudoscalar $A$ to the SM fields are 
\begin{align}
\label{Eq:cf(H)}
c_{H f}& = \sa / \sb \simeq  -\delta - \cotb 
\\
c_{H V} &= \cba \simeq  -\delta
\\
c_{A f} &= i \cb / \sb = i \cot_\beta  \ . 
\end{align}
Therefore to produce $A$ at the level of $\sim$ 5 fb via gluon fusion we have a lower limit on $\cot_\beta^2 \gtrsim 2.5 \times 10^{-3}$ from Eq.~(\ref{Eq:prod}).

\begin{figure}
\includegraphics[width=0.4\textwidth]{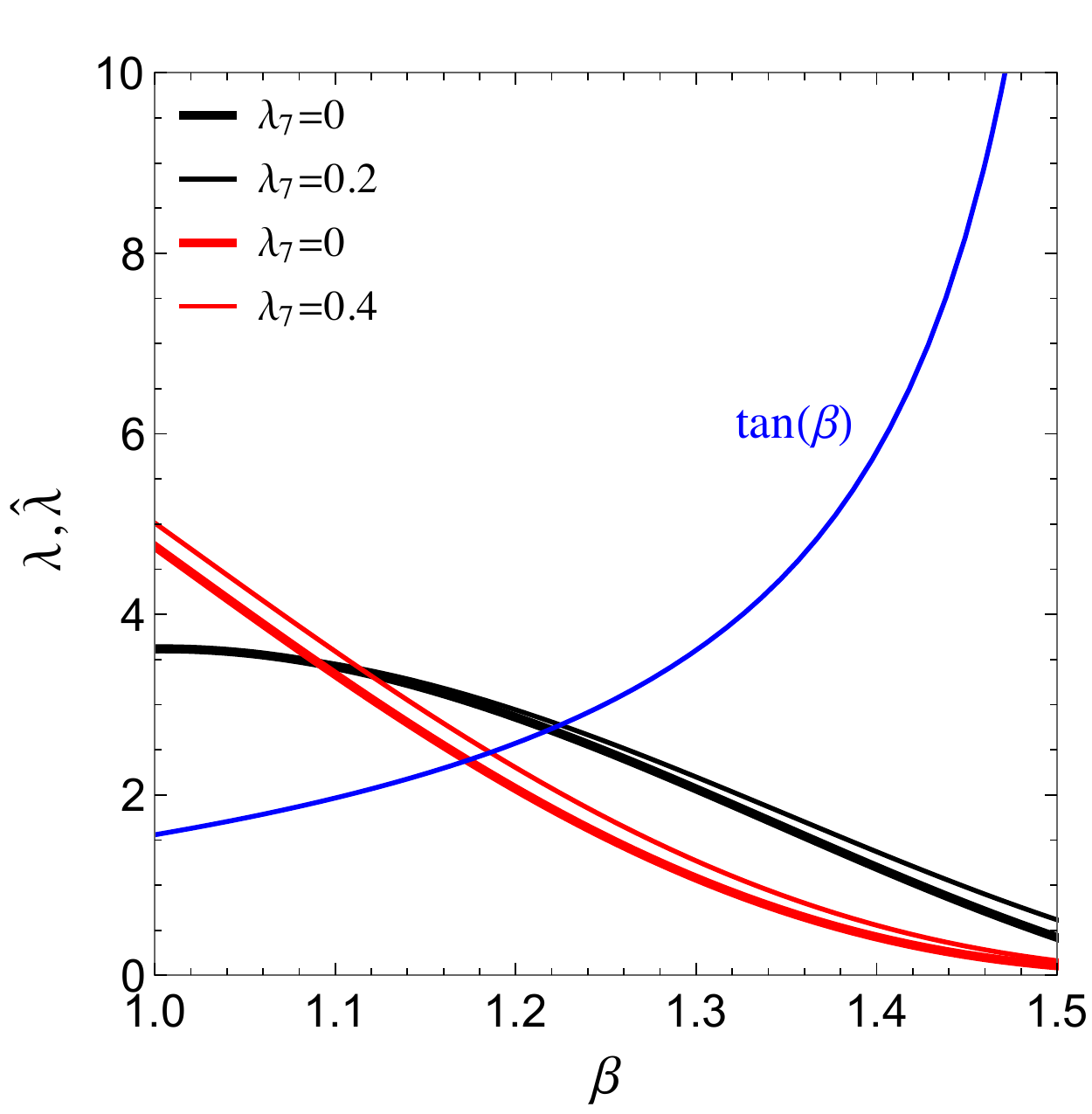}
\includegraphics[width=0.4\textwidth]{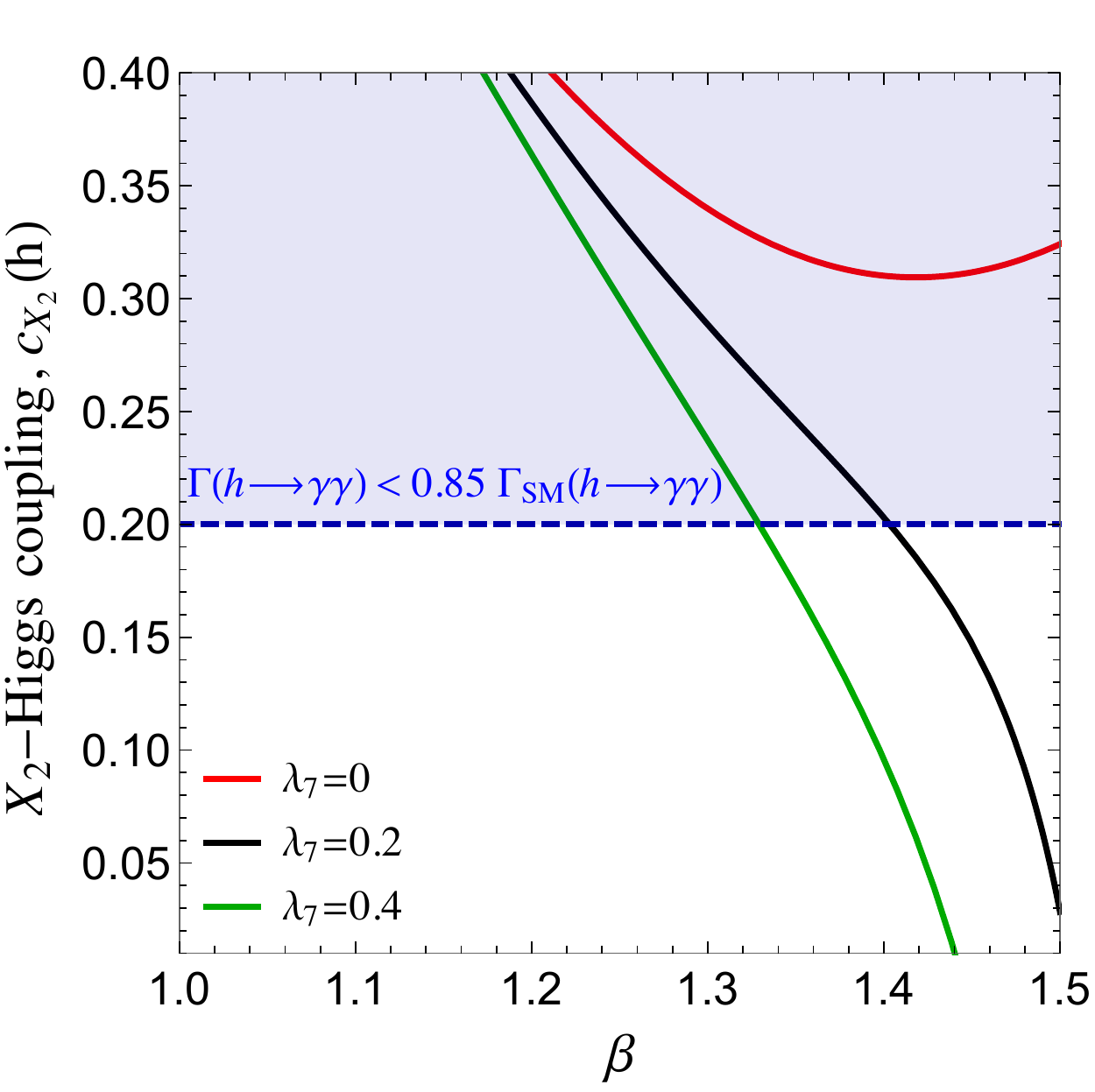}
\caption{Left: The values of $\lambda$, $\widehat{\lambda}$ and $\tan\beta$ as a function of the angle $\beta$.  For $\beta \sim 1.45$ we find values of $\hat{\lambda},\lambda$ corresponding to $m_h\simeq 125$ GeV, $m_A\simeq 750$ GeV and $\tanb \sim O(10)$.
Right: The value of the Yukawa coupling of the heavy fermions to the 125 GeV Higgs as a function of the angle $\beta$. In both plots the additional parameters of the scalar potential are $\lambda_1=0.1,\lambda_2=0.05, \lambda_{345}=6, \lambda_6=2$.}
\label{fig:lambdas}
\end{figure}

Finally the couplings to the new fermions are
\begin{align}
c_{A X_2}& = - i  \tanb 
\\
c_{H X_2} &= \ca / \cb \simeq  \tanb-\delta
\\
c_{h X_2} &= -\sa / \cb \simeq 1+ \delta \, \tanb
\label{Eq:X2couplings}
\end{align}

To avoid large contributions of the new states to the higgs diphoton rate, we require $\delta \, \tan\beta \sim -1$.
With the previous parameters $\delta \sim - 0.1, \hat{\lambda}\simeq1, \lambda=\lambda_A\simeq1/3$ this requires $\tan{\beta}\sim O(10)$. 
In Fig.~\ref{fig:lambdas} on the left panel, we show the value of $\hat{\lambda},\lambda$ and $\tan\beta$ as a function of $\beta$. On the right panel we show 
$c_{h X_2}=1+\delta \, \tan\beta$ for the same parameters. For $\beta \sim 1.45$ we find values of $\hat{\lambda},\lambda$ corresponding to $m_h\simeq 125$ GeV, $m_A\simeq 750$ GeV and $\tanb \sim O(10)$.

With the above couplings we identify our diphoton resonance with the $CP$-odd resonance $\phi\equiv A$ and the diphoton decay rate of $A$ is given as 
\begin{eqnarray}
\Gamma_{A\gamma\gamma} 
\simeq  \frac{\alpha^2 G_F}{32 \sqrt{2}\pi^3} \tan\beta^2 m_A^3
\left|  A^A_{X_2}(\tau_{X_2}) 
\right|^2
\end{eqnarray}

\begin{figure}
\includegraphics[width=0.45\textwidth]{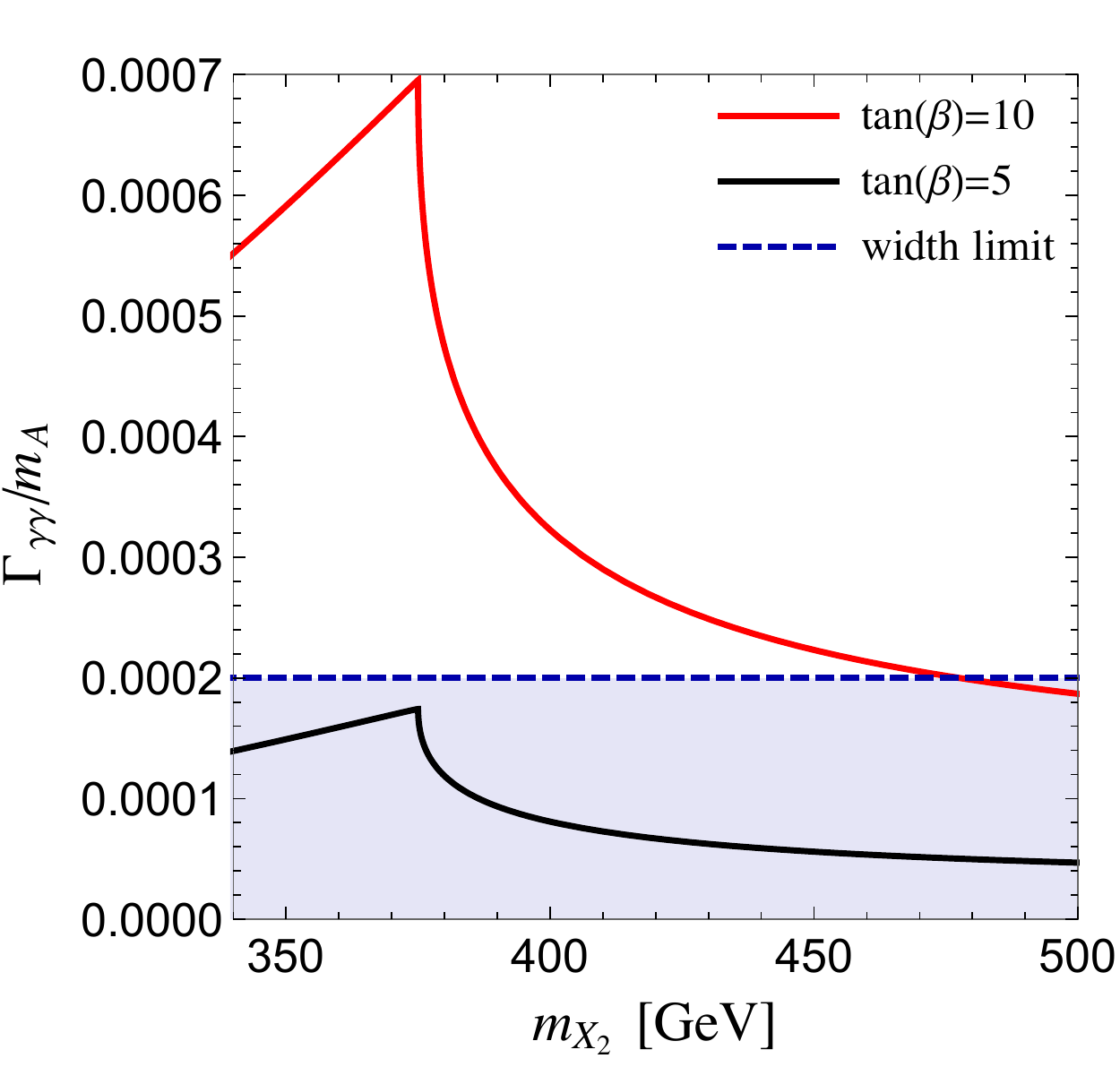}
\caption{The diphoton width of the $CP$-odd resonance $\phi\equiv A$ as a function of the new weak doublet fermion masses $m_{X_2}$ for two different values of $\tan\beta$.}
\label{fig:Gamma}
\end{figure}

In figure~\ref{fig:Gamma} we show the diphoton rate as a function of the fermion masses $m_{X_2}=m_{\widetilde{X}_2}$ near threshold for $\tan\beta=5,10$.
As is clear from the figure we can get sufficient diphoton rate near threshold for $\tan\beta \sim O(10)$.

In particular, using Eq.~\ref{Eq:prod} to express the scaling of the diphoton cross-section we find 
\bea
\sigma_{\gamma\gamma}&\simeq &\left( (5-8) \frac{\tan\beta^2}{10^2} + (1-2)\frac{\tan\beta^6}{10^6}  \right)\times  \frac{1}{1+ 5\times 10^{-5} \tan\beta^4} \, {\rm fb}  \ . 
\label{Eq:diphotonscaling}
\eea
where the range refers to $m_{X_2}=380-400 $GeV and we have assumed $\Gamma_{A}\simeq \Gamma_{A\to tt} + \Gamma_{A\to \gamma\gamma}$.
It follows that for $\tan\beta \sim O(10)$ and near threshold we can also accommodate the diphoton excess. 

\subsection{Constraints}
We conclude the model discussion by briefly summarizing relevant constraints on the model from vacuum stability, Higgs decay into diphotons and searches for weakly charged particles.

Necessary conditions for the stability of the scalar potential in Eq.~(\ref{Eq:scalarpotential}) are \cite{Ferreira:2004yd,Branco:2011iw},
\begin{align}
\lambda_{1,2}& >0 \ , \quad  \lambda_3 > - \sqrt{\lambda_1 \lambda_2} \ , \quad \lambda_3 + \lambda_4 - |\lambda_5|  >  - \sqrt{\lambda_1 \lambda_2}
\ , \quad
2 |\lambda_{6}+ \lambda_{7}| < \frac{\lambda_{1} + \lambda_{2} }{2} + \lambda_{345} Ê.
\label{Eq:stability}
\end{align}
These conditions are satisfied for our representative parameter choices $\lambda_1=0.1,\lambda_2=0.05, \lambda_{345}=6, \lambda_6=2, \lambda_7=0-0.4$ in Fig.~(\ref{fig:lambdas}).  

The production cross-section $\sigma_{h\to\gamma\gamma}$ is however affected by the presence of the new doublets. We define the ratio
$R_{gg}(\gamma\gamma)=\sigma_{h\to\gamma\gamma}/\sigma_{h_{\rm SM}\to\gamma\gamma}$ and expand the reduced couplings $c_h$ as $c_h=1+\delta c_h$ and finally $\delta c_{\gamma\gamma} \mathcal{M}^{\rm SM}_{\gamma\gamma}$ parameterizes new contribution to the diphoton amplitude --- here from the new fermions - in units of the SM ones. We can then write \cite{Alves:2012ez}
\bea
R_{gg}(\gamma\gamma) \simeq 1 + 2 \delta c_{h \gamma\gamma} -0.07 \delta c_{h t} + 2.1 \delta c_{hV} \simeq 1 +  0.8 c_{h X_2} 
\eea
From the right panel of Fig.~(\ref{fig:lambdas}) and the current limit on the diphoton decay width from ATLAS of $\mu=1.17^{+0.27}_{-0.27}$ \cite{Aad:2015gba} and CMS \cite{Khachatryan:2014jba}, we see that in the relevant part of parameter space we are in good agreement with the limit. 

The estimate in Eq.~(\ref{Eq:diphotonscaling}) assumes that the the top and diphoton partial widths dominate the width of $\phi$. This requires the partial decay widths 
$A\to X_2 X_1\to 2 X_1 \, 2\gamma$ and $A\to X_2 X_2^*\to 2 X_1 \to 4 \gamma$ to be negligible. The former requires $y_3 < \cot\beta$ where $\cot\beta \sim (1-2) \times 10^{-1}$ above.  This also immediately suppresses (beyond the off-shell suppression) the second process since $A\to X_2^* X_2^*\to 2 X_1 \to 4 \gamma\sim y_3^4$.

There is still sufficient freedom in the scalar potential to ensure that e.g. $\lambda_A, \lambda_F$ such that $m_H\gtrsim m_A $ and $m_{H^+}\gtrsim m_A$. The $H^0$ state can contribute to the diphoton signal, but the gluon fusion production of a scalar is smaller. Also there is two small canceling contributions in $c_f(H)$ as well as between the W-loop (even if suppressed) and the new fermions in Eq.~(\ref{Eq:cf(H)}) so we expect a subleading contribution to the diphoton signal.

Finally there are no flavor changing neutral currents in the model since only one doublet is coupled to the SM fermions and flavor constraints on the charged $H^\pm$ are very weak for large $\tan\beta$ as considered here  \cite{Mahmoudi:2009zx}. Limits on the new weak charged fermions have been recently summarized in~\cite{Liu:2013gba} and require only that they have masses $\gtrsim 130$ GeV.  While recent ATLAS searches of the charged Higgs states in the $H^{\pm} t ~b$ channel constrain the charged Higgs mass to be $m_{H^{\pm}} \gtrsim 160$ GeV~\cite{Aad:2015typ} based on 20~${\rm fb}^{-1}$ of data at $\sqrt{s} = 8$ TeV.

\section{Summary}
The recently observed excess of diphotons by the ATLAS and CMS experiments may be interpreted as a new spin-0 resonance with a sizable diphoton partial width induced by loops of new EW charged fermions. If these excesses are confirmed with future data it is of great interest to determine how this resonance fits into a broader framework able to address the origin of EWSB and dark matter.

Models of asymmetric DM cogenesis, in which SM sphalerons transfer an initial baryon asymmetry into DM, also require new spin-0 weak doublets coupled to new fermionic weak doublets. In an explicit model example the scalar sector is an extended Type-I 2HDM and we showed that in a part of the parameter space we can accommodate the observed diphoton excess while being in agreement  with current constraints, including the diphoton width of the 125 GeV scalar. Maintaining perturbative Yukawa couplings require the new fermions to be near threshold, around 400 GeV,  and we also commented on a composite realization. 
 
Lastly, we emphasize that low-threshold direct detection offers one of the best avenues to detect the class of dark matter models here (see e.g. the right panel of Fig.~\ref{fig:annihilation}), which prefer DM at the $\mathcal{O}(\rm{GeV})$ scale.

\bibliographystyle{JHEP}

\bibliography{nu}

\end{document}